\documentclass[journal,twoside,web]{IEEEtran}
\usepackage{hyperref}
\usepackage{graphicx}
\usepackage{color}

\title{ECG Feature Importance Rankings: Cardiologists vs.\ Algorithms}
\author{Temesgen Mehari$^{1,2}$, Ashish Sundar$^3$, Alen Bosnjakovic$^4$, Peter Harris$^3$, Steven E. Williams$^5$, Axel Loewe$^6$, Olaf Doessel$^6$, Claudia Nagel$^6$, Nils Strodthoff$^7$, Philip J. Aston$^{3,8}$\thanks{This project (18HLT07 MedalCare) has received funding from the EMPIR programme co-financed by the Participating States and from the European Union’s Horizon 2020 research and innovation programme.}
\thanks{
$^1$ Heinrich Hertz Institute, Berlin, Germany\\
$^2$ Physikalisch-Technische Bundesanstalt, Berlin, Germany\\
$^3$ National Physical Laboratory, Teddington, UK\\
$^4$ Institute of Metrology of Bosnia and Herzegovina, Sarajevo, Bosnia and Herzegovina\\
$^5$ University of Edinburgh, Edinburgh, UK\\
$^6$ Karlsruhe Institute of Technology, Karlsruhe, Germany\\
$^7$ University of Oldenburg, Oldenburg, Germany\\
$^8$ University of Surrey, Guildford, UK
}}

\date{\today}

\begin{document}

\maketitle
\begin{abstract}
Feature importance methods promise to provide a ranking of features according to importance for a given classification task. A wide range of methods exist but their rankings often disagree and they are inherently difficult to evaluate due to a lack of ground truth beyond synthetic datasets. In this work, we put feature importance methods to the test on real-world data in the domain of cardiology, where we try to distinguish three specific pathologies from healthy subjects based on ECG features comparing to features used in cardiologists' decision rules as ground truth. Some methods generally performed well and others performed poorly, while some methods did well on some but not all of the problems considered.
%We find that different methods give good results on different problems and so there is little consistency, but some general trends are identified.
\end{abstract}

\section{Introduction}

A trained cardiologist can diagnose over 150 different conditions from a 12-lead electrocardiogram (ECG) \cite{ECGdiagnosis}. Such diagnoses are made on the basis of a multitude of ECG features which consist mainly of time intervals between certain fiducial points on the ECG, amplitudes of prominent features or morphology of ECG segments. For each pathology, the relevant criteria for specific features are well documented \cite{ECGdiagnosis,chou}, although there may be minor differences between one reference source and another.

On the other hand, there are numerous algorithms available for determining a ranking of features by importance for a given classification task \cite{FI}. However, if several algorithms are used, then it is often found that they give significantly different feature importance rankings and it is not at apparent which ranking is best or whether one particular ranking is better than another. Therefore, we did a comparison of feature importance rankings generated by a number of different algorithms with the corresponding features that a cardiologist uses for diagnosis. This has the advantage of having a set of important features which has been gleaned from clinical experience over many years for diagnosis of each condition which can be compared with the feature rankings of the algorithms.

Another possibility with this study is that the feature importance algorithms could identify features that are important for the diagnosis of a condition which are not normally considered to be important by cardiologists.

We have chosen three pathologies to study, namely first degree atrioventricular block (1$^{\rm st}$ degree AV block), complete right bundle branch block (RBBB) and complete left bundle branch block (LBBB). A diagnosis of these conditions by cardiologists involves 1, 7 and 14 features respectively and so are progressively more complex, starting with the simplest possible case.

For this study, we restrict attention to the simplest case of a binary classification that seeks to distinguish healthy subjects vs.\ a specific pathology. Of course in practice, a cardiologist has to identify a condition (or multiple conditions) out of many possible conditions, which is a much more complicated task. On the other hand, it is quite conceivable that a simple binary classification of healthy vs.\ a specific pathology could be successfully achieved by using only a reduced subset of the complete list of diagnostic conditions. However, we consider it appropriate to study the simplest case first. A study of multiclass feature importance algorithms with all four of the above classes has been undertaken as a separate study \cite{CinC2022}.

We are considering the features used by cardiologists for diagnosis to be the gold standard against which we compare various algorithms. However, it should be noted that different sources for ECG diagnosis often give slightly different conditions for diagnosis of a specific pathology. This may be because textbooks give sufficient conditions for diagnosis, rather than an exhaustive list of all changes associated with a pathology. We have used \textit{EKG-Kurs f\"ur Isabel} \cite{EKG} as it gives simple, itemised conditions for each pathology. More comprehensive texts are available but we chose this one based on its simplicity and clarity.
%We have used the textbook \textit{Chou's Electrocardiography in Clinical Practice} \cite{chou}, which is self-declared to be the ``optimal electrocardiography reference for practicing physicians, and consistently rated as the best choice on the subject for board preparation'', as our reference. However, in our analysis, we will consider features that are not referred to in this text, but which correlate with such features, as these could potentially be ranked highly by the algorithms instead of the features listed in the textbook.

\section{Materials and Methods}
\subsection{ECG Signals}

The ECG signals that were used for this study were taken from the PTB-XL dataset \cite{PTB-XL,PTB-XL-physionet}, which is publicly available on PhysioNet \cite{physionet}. In particular, for each of the three pathologies considered (1$^{\rm st}$ degree AV block, RBBB, LBBB), we extracted all the records that were labelled with only the specific pathology. %\colnst{single-label?}

\subsection{ECG Features}
\label{ECGfeatures_sec}

For extracting features from an ECG, we used the University of Glasgow 12-lead ECG analysis algorithm which has been developed over many years by a team at the University of Glasgow \cite{Glasgow}. This software can derive more than 772 global and lead-dependent ECG features from a 10-second 12-lead ECG signal. (All the features derived by the Glasgow software for the PTB-XL dataset are available in the PTB-XL+ feature dataset \cite{PTBXL+}.) From this large collection of features, we selected 117 which a cardiologist would typically assess when considering a diagnosis that are given in Appendix~\ref{ECGfeatures}. This list of features could be debated, and some might argue for different features to be included, but there is no definitive list of such features. These features were derived for all of the ECG records in each of the pathology classes. The small number of records that contained missing values due to issues with feature extraction were deleted to obtain a final dataset without missing values. Features were similarly extracted from a random selection of an equal number of healthy patients' records, with random replacement of any records containing missing values. With this approach, a balanced dataset containing no missing values was created for each pathology. 

Each feature was scaled to have mean zero and variance one to give a normalised dataset, which was required for certain algorithms (Logistic regression) or is known to be beneficial for others (Deep networks). However, unscaled data were used for XGBoost (XGB) since the feature importance vectors for each test record were almost all zero using the scaled data.

The final datasets contained a total of 1,592 records for 1$^{\rm st}$ degree AV block, 1,074 records for RBBB and 1,072 records for LBBB, with half being for healthy subjects and half for the specific pathology in each case.

\subsection{Pathologies}
% general description of the cardiac conduction system, very basic, just focusing on electrophysiology, leaving out everything related to contraction of the chambers
The ECG is the difference in electrical potential measurable between two different electrodes attached to the body surface and captures the electrical activity due to de- and repolarization of cardiomyocytes in the heart. In the healthy case, electrical activity is spontaneously initiated in the pacemaker cells at the sinoatrial node in the right atrium. After spreading throughout the atrial myocardial tissue and causing the P~wave in the ECG, the excitation is delayed at the atrioventricular node. The electrical activation is then conducted via the bundle of His, which branches into a right bundle as well as an anterior and a posterior left bundle before it reaches the Purkinje fibers. These activate the ventricular myocardium from the apex to the base and lead to the QRS~complex in the ECG. Finally, the T~wave in the ECG arises due to repolarization of the ventricular myocytes. 

% pathologies, what does change
\subsubsection{Atrioventricular Block}
%\colnst{Need a proper reference- EKG fuer Isabel is not acceptable for cardiologists}
In patients with atrioventricular (AV) block, the excitation conduction between atria and ventricles is impaired. In first degree AV block, which is studied in this work, the conduction is markedly delayed and leads to PR intervals $>$200\,ms in the ECG. However, all atrial impulses are still transferred to the ventricles and every P~wave is followed by a QRS complex as opposed to second or third degree AV-block that is associated with skipped beats or independent excitation of atria and ventricles respectively~\cite{EKG}. Thus, there is only one feature which is used for the diagnosis of a 1$^{\rm st}$ degree AV block:
\begin{itemize}
\item 
PR interval
\end{itemize}
We checked for other features that correlate (with absolute Pearson correlation coefficient $\geq 0.7$) with the PR interval, as such features may be expected to occur high up the ranking. However, there were none and so this is the simplest possible case.

\subsubsection{Right Bundle Branch Block}

\begin{table}[t]
    \centering
    \caption{Features that correlate with the important features for RBBB, but not including other important features, together with their correlation coefficients.}
    \begin{tabular}{|l|l|}
    \hline
         & \textbf{Correlating Features} \\
         \textbf{Feature} & \textbf{(correlation coefficient)} \\
    \hline
    \hline
        R amplitude, lead V1 & $-$ \\
        \hline
        R' amplitude, lead V1 & Peak-to-peak amplitude, lead V1 (0.79) \\%0.7934
        \hline
        S amplitude, lead I & $-$ \\ % R' amplitude, lead aVR (-0.6994) \\
        \hline
        S amplitude, lead aVL & $-$ \\ % R amplitude, lead III (-0.6427) \\
        %  & R' amplitude, lead III (-0.6579) \\
        \hline
        S amplitude, lead V1 & $-$ \\ % S amplitude, lead aVR (0.6860) \\
        %  & ST slope, lead V1 (-0.6536) \\
        \hline
        S amplitude, lead V6 & R' amplitude, lead V5 (-0.71) \\%0.7126
        %  & S amplitude, lead II (0.6340) \\
        %  & S amplitude, lead III (0.6388) \\
        %  & S amplitude, lead aVF (0.6630) \\
          & S amplitude, lead V5 (0.82) \\%0.8187
    \hline
       QRS duration & $-$ \\ % ST slope, lead V1 (-0.6405) \\
        %  & QTc Framingham (0.6331) \\
        \hline
    \end{tabular}
    
    \label{tab:RBBB_corr}
\end{table}

Complete right bundle branch block is characterized by marked delay or block in conduction in the right bundle branch. In this case, the right ventricles are activated via impulses conducted through the left bundle branches reaching the right ventricle through the ventricular myocardial tissue. As this takes longer than the physiological activation through the three fascicles, this reflects in a widened QRS~complex of $>$120\,ms in the ECG. Furthermore, a terminal R'~peak %\colpa{what does this mean?} 
is visible in lead V1 and a notched S~wave occurs in leads I, aVL and V6~\cite{EKG}. Thus, the 7 features that are relevant for diagnosis of right bundle branch block are:
\begin{itemize}
\item
QRS duration
\item 
R amplitude in lead V1
\item
R' amplitude in lead V1
\item
S amplitude in leads I, aVL, V1 and V6
\end{itemize}
We call these 7 features the important features for RBBB. We checked for features that correlate (with absolute Pearson correlation coefficient $\geq 0.7$) with one of these 7 features. There were 3 such features, not including the important features above, which are given in Table \ref{tab:RBBB_corr}.

\subsubsection{Left Bundle Branch Block}

\begin{table}[t]
    \centering
    \caption{Features that correlate with the important features for LBBB, but not including other important features, together with their correlation coefficients.}
    \begin{tabular}{|l|l|}
    \hline
         \textbf{Feature} & \textbf{Correlating Features (correlation coefficient)} \\
    \hline
    \hline
    QRS duration & Q amplitude, lead V4 (-0.71) \\
      & S amplitude, lead V3 (-0.71) \\
      & T+ amplitude, lead V1 (0.75) \\
      & ST slope, lead I (-0.73) \\
      & ST slope, lead V1 (0.77) \\
      & ST slope, lead V6 (-0.70) \\
      & ST duration (-0.74) \\
      & T morphology, lead I (-0.82) \\
      & T morphology, lead aVR (0.77) \\
      & T morphology, lead V6 (-0.79) \\
    \hline
    Q amplitude, lead V1 & Peak-to-peak amplitude, lead V1 (-0.90) \\
      & T+ amplitude, lead V1 (-0.79) \\
    \hline
    R amplitude, lead I & Peak-to-peak amplitude, lead I (0.94) \\
    \hline
    R amplitude, lead aVL & Peak-to-peak amplitude, lead I (0.73) \\
      & Peak-to-peak amplitude, lead aVL (0.90) \\
      & Q amplitude, lead III (-0.87) \\
      & Q amplitude, lead aVF (-0.79) \\
      & S amplitude, lead III (-0.81) \\
      & QRS frontal axis (-0.72) \\
    \hline
    R amplitude, lead V5 & Peak-to-peak amplitude, lead V5 (0.86) \\
      & Peak-to-peak amplitude, lead V6 (0.78) \\
      & R amplitude, lead V4 (0.77) \\
    \hline
    R amplitude, lead V6 & Peak-to-peak amplitude, lead V5 (0.72) \\
      & Peak-to-peak amplitude, lead V6 (0.95) \\
    \hline
    R' amplitude, lead I & Peak-to-peak amplitude, lead aVL (0.75) \\
    \hline
    R' amplitude, lead aVL & Peak-to-peak amplitude, lead aVL (0.72) \\
      & S amplitude, lead II (-0.70) \\
      & S amplitude, lead III (-0.78) \\
      & S amplitude, lead aVF (-0.78) \\
    \hline
    R' amplitude, lead V5 & $-$ \\
    \hline
    R' amplitude, lead V6 & $-$ \\
    \hline
    S amplitude, lead I & R' amplitude, lead aVR (-0.71) \\
      & T+ amplitude, lead aVR (-0.77) \\
    \hline
    S amplitude, lead aVL & R amplitude, lead III (-0.74) \\
      & R' amplitude, lead II (-0.74) \\
      & R' amplitude, lead III (-0.72) \\
    \hline
    S amplitude, lead V5 & $-$ \\
    \hline
    S amplitude, lead V6 & R' amplitude, lead V1 (-0.81) \\
    \hline
    \end{tabular}
    \label{tab:LBBB_corr}
\end{table}

Analogously to right bundle branch block described above, complete left bundle branch block describes the condition of a blockage in the electrical conduction in the left bundle branch. As the left bundle branches into an anterior and a posterior fascicle, the term complete left bundle branch block refers to a conduction block before the bifurcation. In the ECG, the delayed activation of the left ventricle reflects in a widened QRS~complex of $>$120\,ms, deep Q~waves in lead V1 and a notched or monophasic QRS morphology in the lateral leads I, aVL, V5 and V6~\cite{EKG}. Thus, there are 14 features that are involved in the diagnosis of left bundle branch block:
\begin{itemize}
\item
QRS duration
\item 
Q amplitude in lead V1
\item
R amplitude in leads I, aVL, V5 and V6
\item
R' amplitude in leads I, aVL, V5 and V6
\item
S amplitude in leads I, aVL, V5 and V6
\end{itemize}
We call these 14 features the important features for LBBB. We checked for features that correlate (with absolute Pearson correlation coefficient $\geq 0.7$) with one of these 14 features, excluding the important features listed above. There were 28 such features that are given in Table \ref{tab:LBBB_corr}.

%\subsubsection{Left ventricular Hypertrophy}
%
%\colnst{I really think we need more content to spice this up if want to get this published at JBHI or TBME}

\subsection{Feature Importance Algorithms}
\label{algorithms}

We can broadly categorize the feature importance algorithms investigated in this work as model-dependent and model-independent methods.

\begin{table}[t]
	\centering
	\caption{Hyperparameters for the machine learning models used.}
	\begin{tabular}{|l|l|}
		\hline
		\textbf{Model} & \textbf{Hyperparameters} \\
		\hline
		\hline
		Random Forests		 & number of trees = 100  \\
							     		&  criterion = `Gini impurity' \\
		\hline
		Boosted Decision Trees      & loss function = `binary logistic'  \\		
		(XGB)  										& learning rate = 0.01 \\ 
													& early stopping rounds = 20 \\
													& number of boosting iterations = 5000 \\						
		\hline
		Logistic Regression 	 & loss = `$l_2$ norm' \\
											& tol = 0.0001 \\
											& max\_iter = 100\\

		\hline
		Deep Neural Networks & 2 hidden fully-connected layers with dim = 256 \\
											& hidden activations =`relu' \\
                                            & output activation = `sigmoid' \\
                                            & optimizer = `adam' \\
											& loss = `binary cross entropy'\\
      %  \hline
      %  Gaussian Processes & \\
		\hline
	\end{tabular}
	\label{tab:hyperparams}
\end{table}

\subsubsection{Model-dependent feature importance methods}
%\colpa{\em We should state which methods are random, and so give different results each time they are run, and which are deterministic.}
\begin{itemize}
\item \textbf{Random forests, Boosted decision trees, Logistic regression and Deep neural networks with permutation/SHAP/LIME feature importance.}
In terms of models, we consider Random forests, Boosted decision trees, Logistic regression and Deep neural networks. The hyperparameters used are summarized in Table \ref{tab:hyperparams}. %\colnst{(Temi: need some details on hyperparameters, TODO: we should provide the code to perform the analysis)}. 
The training data consisted of records from the PTB-XL stratified folds 1--9 and the test data were records from fold 10 \cite{PTB-XL}.
These models were then combined with established attribution methods LIME \cite{ribeiro2016should}, SHAP \cite{NIPS2017_7062} and permutation feature importance \cite{JMLR:v20:18-760}. LIME involves training an interpretable, local surrogate model to approximate the model behaviour near the sample of interest. SHAP is an efficient implementation of the game-theoretic Shapley value approach. 
%Unlike permutation feature importance, which is a global importance measure, 
LIME and SHAP are local attribution methods which return attribution scores per sample, and we therefore ranked the features on the mean of the absolute attribution values across the test set. 
%SHAP considers the mean of the absolute attribution values as a global notion for feature importance. Here, we also consider the mean of the signed attributions. Whereas the former ranks features according to their overall relevance for the prediction problem, the latter identifies features in the order they contribute towards the positive class, which is in this case the interpretation that is more aligned with provided ground truth.
%\colnst{should we include only features with positive attribution values in this case?}
As a third class of feature importance algorithms, we considered permutation feature importance, a global attribution method which quantifies feature importance via the decrease in model performance upon replacing a feature column of interest by a permuted copy of itself. %\colnst{Temi: logistic regression inherently interpretable based on parameter values, can we have permutation feature importance for the other methods, I assume RF permutation feature importance is the impurity-based permutation feature importance which is known to have issues with high cardinality features}. 
%Finally, we include the random forest impurity-based feature ranking.
%\colnst{How about Gaussian processes? Which measure of feature importance? We should apply LIME, SHAP and permutation feature importance as well for GP to make it consistent}
\item \textbf{Random forests.} For a random forest model, the importance of features can be determined by how much they decrease Gini impurity when averaged over all the trees in the forest. It is known that these feature importance values can be misleading for high cardinality features. However, permutation feature importance (see below) can mitigate this to some extent \cite{RF}.
\item \textbf{Logistic regression.} The importance of features in a logistic regression model can be determined by the exponential of the weight associated with each feature \cite{FI}.
\item
\textbf{Gaussian processes.} In Gaussian Process binary classification, the probability of class membership conditioned on an observed feature vector $x$ is modelled as $\sigma(f)$ where $\sigma$ is a sigmoid function, such as the logistic function, and a Gaussian Process model $\textrm{GP}(0,k(x,x'))$ is used as a prior distribution for the latent variable $f$ \cite{GP}. 
%The approach is a generalisation of that of, for example, linear logistic classification in which the latent variable is modelled by a linear function of the features, $f=\mathbf{w}^T\mathbf{x}$, and a multivariate normal is used as the prior distribution for the weights $\mathbf{w}$. 
Using a squared exponential covariance kernel for $k(x,x')$ with diagonal covariance matrix, each feature $x_i$ is associated with its own length-scale parameter $l_i$. A small value for $l_i$ implies the feature varies over short-length scales and so is important for the classification. Consequently, sorting the length-scale parameters provides a ranking of the features. 
%\colnst{should include a standard GP reference here}
\end{itemize}
\subsubsection{Model-independent feature importance methods}
In addition to model-dependent methods we also include methods that solely rely on the data distribution without making use of a trained predictor on the dataset. In the feature selection literature \cite{BolnCanedo2016,Remeseiro2019,BolnCanedo2019}, these methods are often referred to as filter methods.
%\colab{The feature selection process is a process in which a new subset of the features is selected among all others from the original data set. Usually, the mentioned step is performed to make the model more straightforward and reduce the computation time. In addition to this, the features which do not have a high impact on the final prediction are removed from the model.}
More specifically, we consider the following methods:
\begin{itemize}
    \item \textbf{Chi-square test.} The Chi-square test is a statistical hypothesis test that is valid to perform when the test statistic is chi-squared distributed under the null hypothesis. Each feature is tested individually for independence of the response. A small $p$-value is associated with a feature that has dependence on the response, and so is important. Thus, features are ranked by $-\log(p_i)$, where $i$ is the index of the features
    \cite{chi2}.
    \item \textbf{Maximum Relevance - Minimum Redundancy (MRMR).} The MRMR method reduces redundant features while keeping the relevant features for the model, where redundancy and relevance are quantified in terms of mutual information. It is known that many essential features are correlated and redundant and so the MRMR method selects features taking into account the relevance for predicting the outcome variable and the redundancy within the selected features \cite{MRMR1, MRMR2}.
    \item \textbf{Neighbourhood Component Analysis (NCA).} The NCA method selects features by maximizing the prediction accuracy of classification algorithms. The concept of this method is similar to the k-nearest neighbours classification method, only in the NCA method, the reference point is selected randomly not to be the nearest neighbour for the new point \cite{NCA}.
    \item \textbf{ReliefF.} ReliefF calculates a feature score for each feature depending on feature value differences for neighbours which have the same or a different class, which can then be used to rank the features. The ReliefF method estimates the attribute qualities based on how well they can distinguish between instances near them. This method was initially designed to apply to binary classification problems with discrete or numerical features \cite{Kononenko1994}.
    \item \textbf{Modified ROC AUC.} The receiver operating characteristic (ROC) curve consists of a plot of the false positive rate against the true positive rate as a threshold is moved across the distributions for the two classes. The area under the curve (ROC AUC) is a standardised measure of the degree of separation of the two distributions and varies from 0.5 (no discrimination) to 1 (perfect discrimination) \cite{ROC}. 
    
    We note that the positive class has to be specified and that if this is changed from one class to the other, the ROC AUC values range from 0.5 (no discrimination) to 0 (perfect discrimination). Thus, we define   $\mathrm{Modified~ROC~AUC}=\max(\textrm{ROC AUC}, \textrm{1- ROC AUC})$
    %$$\left\{
    %\begin{array}{ll}
    %\mathrm{ROC~AUC} & \mathrm{if~ROC~AUC}\geq 0.5\\
    %1-\mathrm{ROC~AUC} & \mathrm{if~ROC~AUC}< 0.5
    %\end{array}
    %\right.$$
    This ensures that all values are in the range 0.5 to 1.
    
    A feature ranking can be generated for a binary classification problem by generating a distribution for each of the two classes for each feature individually and then finding the Modified ROC AUC for all of these distributions. The features are then ranked by their Modified ROC AUC value from highest to lowest, which ranks the features according to their ability to discriminate the two classes individually. 
    
    We also note that the $\mathrm{ROC~AUC}$ values aid with interpretation of the features, as $\mathrm{ROC~AUC}> 0.5$ implies that the feature increases due to the pathology whereas $\textrm{ROC~AUC}< 0.5$ means that the feature decreases due to the pathology.
\end{itemize}

\subsection{Scoring algorithm}
\label{scoring}
%\colnst{I think it would make most sense to discuss them in the materials section.}
%\colnst{Philip: what about the consistency check via ecgid?}

When comparing a feature ranking generated by one of the algorithms with the important feature set for diagnosis of a specific pathology, we define a score which enables a simple comparison between different methods, assuming that all features in the set of important features for diagnosis have equal importance. As a first step, we choose a value of $n$, which is the number of top features in each ranking that will be considered. We then take a weighted average based on the ranking of each of the top $n$ features that is contained in the important set, where the first feature has a weighting of $n$, the second a weighting of $n-1$ and so on, so that the $n^\mathrm{th}$ feature has a weighting of 1. This weighted average is then normalised to give a score between 0 and 100 (by dividing by $n(n+1)/2/100$), which we round to the nearest integer.

With this scoring system, an important feature in position 1 of the ranking contributes $200/(n+1)$ to the score, whereas an important feature in position $n$ only contributes $200/(n(n+1))$ to the score. For example, taking $n=5$ and assuming that a ranking has the first, second and fourth features in the important set gives a score of $(5+4+2)/15\times 100\approx 73$.

We also consider the ranking of features that are least able to discriminate between the two classes, which can be defined by the features with the lowest modified ROC AUC values. In particular, we consider the two features with the lowest modified ROC AUC values which, for the three pathologies, are as follows:
\begin{itemize}
\item
1$^{\rm st}$ degree AV block: S amplitude, lead I (modified ROC AUC=0.5006); S amplitude, lead V2 (modified ROC AUC=0.5006)
\item
RBBB: R' amplitude, lead V6 (modified ROC AUC=0.5028); R' amplitude, lead I (modified ROC AUC=0.5037)
\item
LBBB: R amplitude, lead I (modified ROC AUC=0.5002); R' amplitude, lead V6 (modified ROC AUC=0.5009)
\end{itemize}
We refer to these as the \textit{non-discriminating features}.

\section{Results}

We consider results of the feature importance ranking algorithms applied to the feature table for each pathology in turn. The model-dependent methods first require training of a machine learning model for the binary classification problem. The accuracy of the five machine learning models for each pathology on the test data are shown in Table \ref{tab:accuracies}. Clearly, these are all very high.

\begin{table}[t]
    \centering
    \caption{The accuracy of the machine learning models on the test data for each distinguishing each pathology from an equally sized set of normal samples.}
    \begin{tabular}{|l|c|c|c|}
        \hline
        & \textbf{1$^{\rm st}$ degree} &&\\
        \textbf{Model} & \textbf{AV block} & \textbf{RBBB} & \textbf{LBBB} \\
        \hline
        \hline
        Random forest & 95.6\% & 99.1\% & 100\% \\
        XGB & 96.8\% & 98.1\% & 100\% \\
        Logistic regression & 95.6\% & 100\% & 100\% \\
        Deep networks & 94.3\% & 100\% & 100\% \\
        Gaussian processes & 97.8\% & 100\% & 100\% \\
        \hline
    \end{tabular}
    \label{tab:accuracies}
\end{table}

\subsection{Atrioventricular Block}

First degree AV block is defined by the PR interval being greater than 200\,ms \cite{ECGdiagnosis}. Thus, there is a single important parameter for diagnosis in this case, namely the PR interval. We therefore expect this feature to occur high up in the rankings.

\begin{figure}[t]
\centering
\includegraphics[width=9.5cm,trim={1cm 0 0 0},clip]{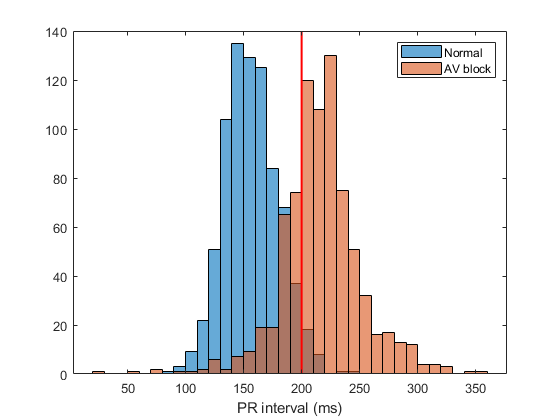}
\caption{Histogram of the PR interval for the records labelled as healthy and 1$^{\rm st}$ degree AV block. The red line is at 200\,ms, which is the threshold for diagnosis of 1$^{\rm st}$ degree AV block.}
\label{PR_interval_AVB}
\end{figure}

For the data we are using, the distributions for the PR interval for the records labelled as Normal and 1$^{\rm st}$ degree AV block are shown in Fig.\ \ref{PR_interval_AVB}. Clearly, not all of the 1$^{\rm st}$ degree AV block records satisfy the diagnostic criterion of exceeding 200\,ms. In fact, the PR interval for 236 out of 796 records labelled as 1$^{\rm st}$ degree AV block does not exceed 200\,ms, with the smallest value being 26\,ms (which is non-physiological). Conversely, there are 23 out of 796 records labelled as Normal that have PR interval exceeding 200\,ms, with the largest value being 242\,ms. Presumably in both cases this is because the Glasgow algorithm identifies the PR interval as shorter/longer than that identified by the cardiologists who labelled the signals.

As an aside, if we tried to classify 1$^{\rm st}$ degree AV block using the Glasgow computed PR intervals, then the Modified ROC AUC for this classification is 0.9384 and the optimal threshold for diagnosis is 184\,ms, which is considerably lower than the conventional 200\,ms threshold. With this threshold, the accuracy of the classification is 88.13\%. Presumably this reduced threshold is as a result of the difference between the PR interval lengths determined by the cardiologists and the Glasgow software.

The ranking of the PR interval by each algorithm is shown in Table \ref{AVB-PR}. These results show that almost all the algorithms we considered ranked the PR interval as the most important feature, although Gaussian processes ranked it second (top feature: T morphology in lead V3) and Deep networks (LIME) ranked it third (top feature: R' amplitude in lead V6). %However, the most notable result is the poor performance of logistic regression and deep networks, which both ranked the PR interval as 19 or below! \colnst{hard to believe- need the feature importance based on parameters to check this} The top ranked feature in these cases was the S amplitude in lead aVR (Logistic regression (LIME)), the S amplitude in lead V2 (Logistic regression (SHAP)), the R' amplitude in lead V1 (Deep networks (LIME)) and the R amplitude in lead II (Deep networks (SHAP)). The relatively low accuracy of Deep networks on the AV block data (see Table \ref{tab:accuracies} could be a possible cause of the poor results for Deep networks, but the same cannot be said for Logistic regression.

\begin{table}[t]
    \centering
    \caption{Ranking of the PR interval and the non-discriminating features when considering Normal and 1$^{\rm st}$ degree AV block signals. Results for model-dependent methods are given in the upper part of the table and results for model-independent methods are given in the lower part.
    %\colnst{for all model-dependent methods permutation feature importance is a simple baseline}
    }
    \begin{tabular}{|l|c|c|}
    \hline
    & & \textbf{Ranking of the} \\
    & \textbf{Ranking of the} & \textbf{non-discriminating} \\
    \textbf{Method} & \textbf{PR interval} & \textbf{features} \\
    \hline
    \hline
    Random forest (permutation) & \textbf{1} & 2, 33 \\
    Random forest (SHAP) & \textbf{1} & 65, 29 \\
    Random forest (LIME) & \textbf{1} & 71, 78 \\
    Random forest & \textbf{1} & 74, 27 \\
    XGB (permutation) & \textbf{1} & 25, 15 \\
    XGB (SHAP) & \textbf{1} & 88, 24 \\
    XGB (LIME) & \textbf{1} & 110, 81 \\
    Logistic regression (permutation) & \textbf{1} & 94, 29 \\
    Logistic regression (SHAP) & \textbf{1} & 113, 55 \\
    Logistic regression (LIME) & \textbf{1} & 76, 85 \\
    Logistic regression & \textbf{1} & 114, 62 \\
    Deep networks (permutation) & \textbf{1} & 52, 74 \\
    Deep networks (SHAP) & \textbf{1} & 95, 82 \\
    Deep networks (LIME) & 3 & 51, 99 \\
    Gaussian processes & 2 & 94, 80 \\
    \hline
    \hline
    Chi-square test & \textbf{1} & 92, 26 \\
    MRMR & \textbf{1} & 97, 99 \\
    NCA & \textbf{1} & 110, 32 \\
    ReliefF & \textbf{1} & 64, 39 \\
    Modified ROC AUC & \textbf{1} & 116, 117 \\
    \hline
    \end{tabular}
    \label{AVB-PR}
\end{table}

We also considered the ranking for each method of the non-discriminating features, which have both got Modified ROC AUC values very close to 0.5. These are shown in the final column of Table \ref{AVB-PR}.  We note that they are by definition the last two features in the Modified ROC AUC ranking. It is surprising that one of the two non-discriminating features is ranked as 2 for Random forest (permutation). The rankings of 25 and 15 for XGB (permutation) is also relatively high. On the other hand, XGB (LIME), Logistic regression (SHAP), Logistic regression and NCA ranked at least one of the two non-discriminating features as more than 100.

We then found the top 5 features for each of the methods to see if there is any commonality between them. The frequency of features in the top 5 is shown in Table \ref{AVB-top5-freq} which, as expecte, includes the PR interval as the most common. Two methods had their top 5 features matching those in Table \ref{AVB-top5-freq} which were Random forest and Random forest (SHAP), while Random forest (LIME), XGB (SHAP) and Chi-square test all had 4 out of these 5 in their top 5. On the other hand, Random forest (permutation), Logistic regression (LIME and SHAP), Logistic regression, Deep networks (LIME), Gaussian processes and MRMR only had the PR interval of those listed in Table \ref{AVB-top5-freq} in their top 5 features. 
%\colnst{some clinical/modeling perspective on this list would be desirable}

The ROC AUC values in Table \ref{AVB-top5-freq} indicate the direction of change of a feature with the pathology as described in Section \ref{algorithms}. Clearly, in this case, the PR interval increases with 1$^{\rm st}$ degree AV block, which is consistent with the cardiologists' diagnosis. 

The QRS duration generally increases with 1$^{\rm st}$ degree AV block. The mean QRS duration for normal subjects is 92\,ms which increases to 113\,ms for 1$^{\rm st}$ degree AV block subjects. This is consistent with evidence of conduction slowing distal to the AV node in patients with known 1$^{\rm st}$ degree AV block.
%This is possibly rather counter-intuitive as it might be expected that sodium channel availability is higher after a longer diastolic interval (due to a longer PR interval), and thus conduction is faster and QRS duration is shorter. 

The T+ amplitude in leads I and V6 decreases on average in patients with 1$^{\rm st}$ degree AV block according to these results. The physiological cause for these decreases is not clear.

Finally, the T morphology measure in lead I decreases with 1$^{\rm st}$ degree AV block, but this is an integer value representing different cases. Analysis of this feature shows that 99\% of the values for the Normal category are +1, indicating a single upright T wave. However, for the 1$^{\rm st}$ degree AV block records, only 52\% have a value of +1, with almost all the others having a value of either $-1$ or $-2$ in equal proportions. Thus, it seems that in approximately half the cases of 1$^{\rm st}$ degree AV block, the T wave is inverted or biphasic with negative leading component. A possible explanation for this is that for 1$^{\rm st}$ degree AV block subjects, the PR interval is longer resulting in a longer diastolic interval. If the action potential duration increases more in some regions than others for longer diastolic intervals (restitution), this could cause morphology changes in the T wave.

\begin{table}[t]
    \centering
    \caption{The most common features in the top 5 for 1$^{\rm st}$ degree AV block for all 20 methods and their Modified ROC AUC and ROC AUC values.}
    \begin{tabular}{|l|c|c|c|}
    \hline
    & \textbf{Frequency in the} & \textbf{Modified} & \\
        \textbf{Feature}  & \textbf{top 5 features} & \textbf{ROC AUC} & \textbf{ROC AUC} \\
    \hline
    \hline
    PR interval & 20 & 0.9384 & 0.9384 \\
    QRS duration & 8 & 0.7450 & 0.7450 \\
    T+ amplitude, lead I & 7 & 0.8247 & 0.1753 \\
    T+ amplitude, lead V6 & 7 & 0.8175 & 0.1825 \\
    T morphology, lead I & 6 & 0.7289 & 0.2711 \\
    \hline
    \end{tabular}
    \label{AVB-top5-freq}
\end{table}

\subsection{Right Bundle Branch Block}

For RBBB, there are 7 important features and a further 3 features that correlate with at least one of these, as listed in Table \ref{tab:RBBB_corr}. Using the scoring algorithm described in Section \ref{scoring} we found the score for each method using the top 5 features of each ranking only. In Table \ref{tab:RBBB}, scores for each method comparing the top 5 features with both the important features and the important and correlating features are given. The best performing method is Logistic regression, while Random forest (SHAP), Random forest, Logistic regression (SHAP and LIME), Deep networks (SHAP), Chi-square test and Modified ROC AUC all have scores over 70. Only Deep networks (permutation and LIME) have an increased score when including the correlating features, but in both cases the score for the important features only is zero. The worst performing methods are Random forest (permutation), Deep networks (permutation and LIME), MRMR and NCA. We also note that SHAP outperformed LIME for each of the four methods.

\begin{table}[t]
    \centering
    \caption{Right bundle branch block top 5 scores for the different feature importance rankings using the important features only or the important features together with features that correlate with them. The ranking of the non-discriminating features is also given. Results for model-dependent methods are given in the upper part of the table and results for model-independent methods are given in the lower part.}
    \begin{tabular}{|l|c|c|}
    \hline
       & \textbf{Top 5 score} & \textbf{Ranking of the} \\
       & \textbf{important/} & \textbf{non-discriminating} \\
           \textbf{Method} & \textbf{imp. + corr.} & \textbf{features} \\
    \hline
    \hline
    Random forest (permutation) & 33/33 & 21, 31 \\
    Random forest (SHAP) & 73/73 & 114, 115 \\
    Random forest (LIME) & 67/67 & 23, 14 \\
    Random forest & 80/80 & 116, 113 \\
    XGB (permutation) & 60/60 & 20, 61 \\
    XGB (SHAP) & 67/67 & 111, 103 \\
    XGB (LIME) & 53/53 & 13, 18 \\
    Logistic regression (permutation) & 60/60 & 24, 63 \\
    Logistic regression (SHAP) & 80/80 & 100, 117 \\
    Logistic regression (LIME) & 73/73 & 14, 12 \\
    Logistic regression & \textbf{93}/\textbf{93} & 78, 114 \\
    Deep networks (permutation) & 0/27 & 87, 112 \\
    Deep networks (SHAP) & 73/73 & 116, 117 \\
    Deep networks (LIME) & 0/13 & 5, 2 \\
    Gaussian processes & 60/60 & 12, 10 \\
    \hline
    \hline
    Chi-square test & 73/73 & 115, 112 \\
    MRMR & 33/33 & 2, 23 \\
    NCA & 13/13 & 53, 58 \\
    ReliefF & 60/60 & 108, 107 \\
    Modified ROC AUC & 73/73 & 116, 117 \\
    \hline
    \end{tabular}
    \label{tab:RBBB}
\end{table}

We then considered the ranking for each method of the non-discriminating features which are shown in the final column of Table \ref{AVB-PR}. We note that these feature rankings are very low for Random Forest (SHAP), Random forest, XGB (SHAP), Logistic regression (SHAP), Deep networks (SHAP), Chi-square test and ReliefF so all the SHAP methods do very well. However, these feature rankings are high for Random forest (LIME), XGB (LIME), Logistic regression (LIME), Deep networks (LIME), Gaussian processes and MRMR so all the LIME methods perform poorly for these features.

\begin{table*}[h]
    \centering
    \caption{The most common features in the top 5 for RBBB for all 20 methods and their Modified ROC AUC and ROC AUC values.}
    \begin{tabular}{|l|c|c|c|c|}
    \hline
     & \textbf{Frequency in the} & \textbf{Type of} & \textbf{Modified} & \\
     \textbf{Feature}  & \textbf{top 5 features} & \textbf{feature} & \textbf{ROC AUC} & \textbf{ROC AUC} \\
    \hline
    \hline
    QRS duration & 18 & Important & 0.9933 & 0.9933 \\
    S amplitude, lead V1 & 12 & Important & 0.9283 & 0.9283 \\
    R' amplitude, lead V1 & 9 & Important & 0.7860 & 0.7860 \\
    ST slope, lead V1 & 8 & Unimportant & 0.9611 & 0.0389 \\
    S amplitude, lead I & 7 & Important & 0.9234 & 0.0766 \\
    S amplitude, lead V2 & 7 & Unimportant & 0.9199 & 0.9199 \\
    \hline
    \end{tabular}
    \label{RBBB-top5-freq}
\end{table*}

We again considered the top 5 features for each method with the 6 most frequent shown in Table \ref{RBBB-top5-freq}. We note that 4 of these are important features, but that ST slope in lead V1 and S amplitude in lead V2 are not, but both have very high Modified ROC AUC values and are positioned as second and sixth respectively in the ROC AUC ranking. Clearly leads V1 and V2 are very close on the body, so it is perhaps not surprising that the S amplitude in lead V2 has significance as well as the important feature of the S amplitude in lead V1. The correlation coefficient between the two is reasonably high at 0.6707. The ROC AUC value for the S amplitude in lead V1 indicates that it increases with RBBB, resulting in a shallower S wave (since the S wave amplitudes are negative) and so it is also not surprising that the ST slope in lead V1 decreases with RBBB, again as shown by the ROC AUC value. The correlation coefficient between the two of -0.6536 is again reasonably high in magnitude.
%\colnst{again a clinical/modeling perspecive would be helpful} 
All these features have a very high Modified ROC AUC value, which indicates good separation of the two distributions for these features, except for R' amplitude in lead V1.

Two methods had all of their top 5 features in Table \ref{RBBB-top5-freq}, namely Random forest (SHAP) and Deep networks (SHAP) while Random forest (LIME), Logistic regression (SHAP), Logistic regression (permutation), Chi-square test, ReliefF and Modified ROC AUC all had 4 of their top 5 features in Table \ref{RBBB-top5-freq}. The worst performing methods were Deep networks (LIME) and Deep networks (permutation) which had no features in Table \ref{RBBB-top5-freq} in their top 5, and NCA and MRMR which both had one feature from Table \ref{RBBB-top5-freq} in their top 5, which was QRS duration in both cases.

The ROC AUC values in Table \ref{RBBB-top5-freq} show that QRS duration increases with RBBB, which is consistent with one of the diagnosis conditions that the width of the QRS complex should be $>$120\,ms. The S amplitude in leads V1 and V2 increases with RBBB, resulting in shallower S waves since the S amplitude is negative,  while the S amplitude in lead I decreases with RBBB, resulting in a deeper S wave. The R' amplitude in lead V1 increases with RBBB. Finally, the ST slope in lead V1 decreases with RBBB.
%which is consistent with a shallower S peak in lead V1.
%The diagnosis conditions above state that RBBB causes a notched S wave in leads V6, which is consistent with a decrease in amplitude. There is no indication in the diagnostic criteria above that the S amplitude increases in leads V1 and V2. However, in the more comprehensive text \cite{chou}, the diagnostic criteria for RBBB include an rsr0, rsR0, or rSR0 pattern in lead V1 or V2. [{\em Could this explain an increase in S amplitude in leads V1 and V2?}] 
%\colnst{again clinical/modeling}

\subsection{Left Bundle Branch Block}

For LBBB, there are 14 important features and an additional 28 correlating features, as listed in Table \ref{tab:LBBB_corr}. The scoring algorithm described in Section \ref{scoring} gives the scores as shown in Table \ref{tab:LBBB}, again using only the top 5 features. The scores for the important features only are generally quite low. However, when the correlating features are included, most methods show a significant improvement, which is not surprising as there are 28 additional correlating features, although much of the improvement in scores is due to the three T morphology features (see Table \ref{LBBB-top5-freq}).

Using only the important features, the best performing method is Gaussian processes, while Logistic regression (permutation) and Deep networks (permutation) both have a score of 0. When the correlating features are included, a perfect score of 100 is obtained by Random forest (SHAP), Random forest, Deep networks (SHAP) and Chi-square test. Logistic regression (permutation) still has a zero score while Deep networks (permutation) has an increased, but still poor, score of 33.

\begin{table}[t]
    \centering
    \caption{Left bundle branch block top 5 scores for the different feature importance rankings using the important features only or the important features together with features that correlate with them. The ranking of the non-discriminating features is also given. Results for model-dependent methods are given in the upper part of the table and results for model-independent methods are given in the lower part.}
    \begin{tabular}{|l|c|c|}
    \hline
       & \textbf{Top 5 score} & \textbf{Ranking of the} \\
       & \textbf{important/} & \textbf{non-discriminating} \\
       \textbf{Method} & \textbf{imp. + corr.} & \textbf{features} \\
    \hline
    \hline
    Random forest (permutation) & 33/40 & 61, 56 \\
    Random forest (SHAP) & 33/\textbf{100} & 107, 63 \\
    Random forest (LIME) & 33/80 & 22, 83 \\
    Random forest & 33/\textbf{100} & 117, 69 \\
    XGB (permutation) & 33/60 & 93, 57 \\
    XGB (SHAP) & 33/73 & 92, 67 \\
    XGB (LIME) & 33/53 & 7, 69 \\
    Logistic regression (permutation) & 0/0 & 75, 56 \\
    Logistic regression (SHAP) & 33/67 & 115, 59 \\
    Logistic regression (LIME) & 27/53 & 2, 85 \\
    Logistic regression & 33/60 & 55, 59 \\
    Deep networks (permutation) & 0/33 & 90, 56 \\
    Deep networks (SHAP) & 33/\textbf{100} & 114, 56 \\
    Deep networks (LIME) & 47/73 & 1, 95 \\
    Gaussian processes & \textbf{60}/87 & 117, 97 \\
    \hline
    \hline
    Chi-square test & 33/\textbf{100} & 116, 109 \\
    MRMR & 33/67 & 113, 117 \\
    NCA & 53/73 & 64, 35 \\
    ReliefF & 13/93 & 95, 71 \\
    Modified ROC AUC & 33/80 & 116, 117 \\
    \hline
    \end{tabular}
    \label{tab:LBBB}
\end{table}
%\clearpage

The rankings of the non-discriminating features were generally low, with Chi-square test and MRMR performing particularly well. However, the various methods combined with LIME gave quite high rankings for one of these features which had rank 1, 2, 7 and 22 for Deep networks, Logistic regression, XGB and Random forest respectively, which is very poor. In contrast, the SHAP methods all ranked this feature greater than 110, except for XGB (SHAP) which ranked it as 92, so these methods all performed well.

The frequency of features in the top 5 for all methods is shown in Table \ref{LBBB-top5-freq}. We note that four of these are correlating features, which explains the big increase in scores when the correlating features are included. Again, all of these 5 features have a very high Modified ROC AUC value, indicating good separation of the two distributions for these features. We note that the four methods that did not have QRS duration in their top 5 features were Logistic regression (permutation and LIME) and Deep networks (permutation and LIME).

No method had all the top 5 features matching those in Table \ref{LBBB-top5-freq} but Deep networks (SHAP), Chi-square test and ReliefF both had 4 out of their top 5 that matched with Table \ref{LBBB-top5-freq}. On the other hand, Logistic regression (permutation) and Deep networks (permutation and LIME) had none of the features in Table \ref{LBBB-top5-freq} in their top 5.

\begin{table*}[t]
    \centering
        \caption{The most common features in the top 5 for LBBB for all 20 methods and their Modified ROC AUC and ROC AUC values.}
    \begin{tabular}{|l|c|c|c|c|}
    \hline
     & \textbf{Frequency in the} & \textbf{Type of} & \textbf{Modified} & \\
     \textbf{Feature}  & \textbf{top 5 features} & \textbf{feature} & \textbf{ROC AUC} & \textbf{ROC AUC} \\
    \hline
    \hline
    QRS duration & 16 & Important & 0.9960 & 0.9960 \\
    T morphology, lead I & 11 & Correlating & 0.9689 & 0.0311 \\
    T morphology, lead V6 & 9 & Correlating & 0.9510 & 0.0490 \\
    T morphology, lead aVR & 4 & Correlating & 0.9300 & 0.9300 \\
    R amplitude, lead V4 & 4 & Correlating & 0.9401 & 0.0599 \\
    \hline
    \end{tabular}
    \label{LBBB-top5-freq}
\end{table*}

The ROC AUC values show that the QRS duration increases with LBBB, which is consistent with the condition that the width of the QRS complex should be $>$120\,ms. The diagnosis of LBBB involves only changes in the QRS complex but the three T morphology features in Table \ref{LBBB-top5-freq} are not associated with the QRS complex. However, we have already noted they correlate strongly with the QRS duration.

The T morphology features for leads I and V6 decrease with LBBB. Analysis of these features shows that 99\% of the values for the Normal class are +1 for both morphology features. For the LBBB records, 72\% are $-1$ and 24\% are $-2$ for the T morphology in lead I, and 69\% are $-1$ and 24\% are $-2$ for the T morphology in lead V6, both of which represent a significant shift from a single upright wave to either a single inverted wave or a biphasic wave with leading negative component. The T morphology in lead aVR increases with LBBB. In this case, 99\% of the values for the Normal class are $-1$ while for the LBBB records, 50\% are +1 and 37\% are +2. So almost all of the Normal class have a single inverted T wave which changes to either a single upright wave or a biphasic T wave with leading positive component.

The R amplitude in lead V4 is not an important feature for the diagnosis of LBBB, but this amplitude in leads V5 and V6 are important features. As lead V4 is very close to lead V5, it is not too surprising that this feature is common in the top 5 features for some methods. Interestingly, the R amplitude in leads V5 and V6 are not in the top 5 features for any method, so lead V4 seems to be more important than leads V5 and V6.

\section{Comparison with the multiclass case}

We have considered feature importance ranking in the context of a binary classification of normal vs. a single pathology for three different pathologies, namely 1$^{\rm st}$ degree AV block, RBBB and LBBB. This is the simplest possible case, but is not very realistic since cardiologists have to positively diagnose one (or more) conditions from a long list of possible conditions. It is also conceivable that a simple binary classification of normal vs.\ a specific pathology could be achieved with high accuracy using only a subset of the complete list of diagnostic conditions. Thus, as a next step, we considered feature importance ranking for a multiclass classification involving normal, 1$^{\rm st}$ degree AV block, RBBB and LBBB records in \cite{CinC2022}. The feature importance rankings were found for the one vs. all binary classifications as the aim is to positively diagnose one condition (since the data were single label) which implies a negative classification for the other classes. 

The accuracies of the models were not reported in \cite{CinC2022} but all four methods had an accuracy exceeding 95\% for the multiclass classification. Also, the results for the model dependent methods are not directly comparable since the data were not normalised in \cite{CinC2022} as they were in this study. In particular, the poor performance of Deep networks for the ranking of the PR interval for the 1$^{\rm st}$ degree AV block case is almost certainly due to this lack of normalisation.

We now compare the feature rankings of the binary and multiclass cases.

\subsection{First Degree AV Block}

The ranking of the PR interval was very similar in the binary and multiclass cases. In the binary case, all methods ranked the PR interval as most important except for Deep networks (LIME) and Gaussian processes, which ranked it as third and second respectively. In the multiclass case, the PR interval was not the top feature for Logistic regression (SHAP and LIME), Deep networks (SHAP and LIME) and Gaussian processes. The poor results for Logistic regression and Deep networks are probably due to the fact that the data were not normalised. We note that although Gaussian processes ranked the PR interval as second in both cases, the top feature differs. For the binary case, the top feature was T morphology in lead V3 while in the multiclass case, the top feature was QRS duration.

The most common features in the top 5 had three features in common, namely the PR interval, QRS duration and T+ amplitude in lead I. The other features listed in Table \ref{AVB-top5-freq} are the T+ amplitude in lead V6 and T morphology in lead I whereas the other features for the multiclass case were the ST slope in leads I and V1 which are quite different features for the two cases.

\subsection{RBBB}

We first note that the correlating features for RBBB were different for the binary and multiclass cases, with 3 correlating features in the binary case (which are listed in Table \ref{tab:RBBB_corr}) and 5 correlating features for the multiclass case. The scores for the important and correlating features for the multiclass case are greater than the corresponding scores for the binary case for many methods, although a notable exception is Logistic regression (SHAP and LIME) which both had a score of zero in the multiclass case and the best score in the binary case! We also note that in the multiclass case, the scores for the important and correlating features were 100 for four methods, namely Random forest, Random forest (permutation) and XGB (SHAP and LIME). The scores for MRMR and NCA were very low for the binary case, but improved significantly for the multiclass case, for which they had the second best score (important features only).

In this case, the 5 most common features in the top 5 in the multiclass case are all included in the 6 most common features in the top 5 for the binary case, but also include the S amplitude in lead V1 as the extra feature. 

\subsection{LBBB}

In this case, there are 28 correlating features in the binary case (which are given in Table \ref{tab:LBBB_corr}) but only 17 correlating features for the multiclass case. The scores for the important features only and for the important plus correlating features for the multiclass case were almost all less than the corresponding scores for the binary case.

The most common features in the top 5 only had no features in common in this case. The multiclass case includes the ST slope in three leads whereas the binary case includes the T moprhology in three leads.

\section{Discussion}

The results of the different feature ranking algorithms for the three pathologies that we have considered have some inconsistencies, although some general trends can be observed. For 1$^{\rm st}$ degree AV block, all methods ranked the one important feature first, except for Deep networks (LIME) and Gaussian processes, which ranked it as third and second respectively. For RBBB, Logistic regression had the highest scores but scored quite poorly for LBBB. For LBBB, a score of 100 when including correlating features was obtained by Random forest (SHAP), Random forest, Deep networks (SHAP) and Chi-square test. Also, NCA scored very poorly for RBBB but did quite well for LBBB. Conversely, ReliefF performed poorly for LBBB (important features only) but had reasonable performance for RBBB. But MRMR performed poorly for both RBBB and LBBB.

%Methods that performed reasonably well for both RBBB and LBBB (including correlating features) include Random forest (permutation and SHAP), Logistic regression (SHAP and LIME), Deep networks (SHAP), Gaussian processes and ROC AUC.

If the scores for RBBB and LBBB are added together, then for the important features only, Logistic regression has the highest score, closely followed by Gaussian processes, Random forest and Logistic regression (SHAP). At the other end, Deep networks (permutation) has a combined score of zero, while Deep networks (LIME) has the lowest non-zero combined score. Adding the scores for RBBB and LBBB for the important and correlating features, then the top score is obtained by Random forest, with Random forest (SHAP), Deep networks (SHAP) and Chi-square test all tied in second place, while the lowest combined score was obtained for Logistic regression (permutation) together with Deep networks (permutation).

When comparing the various methods combined with SHAP, LIME and permutation options, the permutation variations were consistently the worst, followed by LIME, with the best results obtained by SHAP. However, Random forest results were always the same as or better than Random forest (SHAP) and Logistic regression results were the same as or better than Logistic regression (SHAP) except for LBBB including correlating features. So the native feature importance rankings for Random forest and Logistic regression seem to do well without the addition of other methods on top.

All of the SHAP methods together with Chi-square test and Random forest all ranked the non-discriminating features quite far down the rankings for RBBB and LBBB but the LIME methods all put the non-discriminating features quite high up in the rankings. MRMR ranking of the non-discriminating features was particular good for LBBB but was particularly bad for RBBB, and so there is inconsistency here.

For 1$^{\rm st}$ degree AV block, which is diagnosed using the single feature of PR interval, other commonly highly ranked features include QRS duration, as well as T+ amplitude and T morphology in lead I.

For RBBB, two unimportant features were commonly highly ranked namely ST slope in lead V1 and S amplitude in lead V2. It is interesting to note that both of these were correlating features in the multiclass case. These results also suggest that there are significant changes in the S amplitude in lead V2, as well as in V1, that may be considered.

The most surprising results from this work concern LBBB, where the most common features in the top 5 include three T morphology features, in leads I, aVR and V6, whereas LBBB is diagnosed using only features related to the QRS complex (although all three of these features correlate strongly with QRS duration). QRS morphology in leads I and V6 is used in the diagnosis of LBBB, so it is likely that these changes also result in the leading component 
 of the T wave of leads I and V6 changing from positive to negative for 96\% and 93\% of the LBBB records respectively. But lead aVR is not used in the diagnosis of LBBB at all, but is in the top 5 features for 4 of the methods and here the leading component of the T wave changes from negative to positive for 87\% of the LBBB records. Similarly, lead V4 is not used to diagnose LBBB, but the R amplitude in lead V4 also occurs in the top 5 features for 4 of the methods.

\section{Conclusion}

In this comparison of feature ranking algorithms with the expert knowledge of cardiologists for three different pathologies, we have shown that, generally speaking, the SHAP methods all give good agreement with the important features used by cardiologists, together with the native Random forest and Logistic regression feature rankings. For the model independent methods, Chi-square test generally performed well. Some methods gave inconsistent results, including MRMR and NCA. The permutation methods generally performed quite poorly.

It is interesting that the top ranked features for many methods include some unimportant or correlating features rather than important features only.

The code for obtaining the feature importance rankings described in this work is available at \url{https://github.com/tmehari/feature_importance}.
%The main conclusion from this work is that different methods perform well for different problems when comparing feature rankings with features that cardiologists use for diagnosis. The best performing method for one problem can be the worst performing for another. So there is little consistency. However, some general trends can be observed. For model based methods, Random forest methods generally give good scores, as do Gaussian processes while Logistic regression and Deep networks are highly variable. For model independent methods, Chi-square test and Modified ROC AUC generally give good results whereas MRMR, NCA and ReliefF are inconsistent, sometimes giving good results and sometimes bad.

%\begin{itemize}
%    \item 
%    Some features have lots of zero entries. It is not clear how this would affect the feature importance ranking algorithms.
%    \item
%    The morphology features can only take values $\pm 1$ and $\pm 2$.
%    \item
%    T wave morphology parameters important for LBBB but not mentioned in the diagnosis conditions.
%    \item
%    Results from random forests are slightly different each time! Is this the case with other methods as well?
%\end{itemize}

\bibliographystyle{IEEEtran}
\bibliography{bibfile.bib}

\appendices
\section{ECG Features}
\label{ECGfeatures}

The 117 features from the Glasgow 12-lead ECG analysis algorithm \cite{Glasgow} that we identified as ones that cardiologists would typically consider when making a diagnosis mainly consist of features derived for all 12 leads, which are as follows:
\begin{itemize}
\item 
Peak-to-peak amplitude
\item
Q amplitude
\item
R amplitude
\item
S amplitude
\item
R' amplitude (amplitude of a second R peak)
\item
T+ amplitude (maximum height of the T wave)
\item
P morphology
\item
T morphology
\item
ST slope
\end{itemize}
The morphology parameters are integers representing four cases, namely:
\begin{itemize}
\item
A biphasic wave with leading positive component (+2)
\item
A single upright wave (+1)
\item
A single inverted wave ($-1$)
\item
A biphasic wave with leading negative component ($-2$)
\end{itemize}
In addition, a number of measurements derived from all 12 leads were used as follows:
\begin{itemize}
\item
QRS frontal axis
\item
Average RR interval
\item
Heart rate variability
\item
Overall ST duration
\item
Overall PR interval
\item
QTc (Framingham)
\item
Overall P duration
\item
Overall QRS duration
\item
Overall T duration
\end{itemize}
\end{document}